Christopher K. Allsup, Irene S. Gabashvili

Aurametrix, USA

https://aurametrix.com

# Democratizing Strategic Planning in Master-Planned Communities

## Abstract

This paper introduces a strategic planning tool for master-planned communities designed specifically to quantify residents' subjective preferences about large investments in amenities and infrastructure projects. Drawing on data obtained from brief on-line surveys, the tool ranks alternative plans by considering the aggregate anticipated utilization of each proposed amenity and cost sensitivity to it (or risk sensitivity for infrastructure plans). In addition, the tool estimates the percentage of households that favor the preferred plan and predicts whether residents would actually be willing to fund the project. The mathematical underpinnings of the tool are borrowed from utility theory, incorporating exponential functions to model diminishing marginal returns on quality, cost, and risk mitigation.

## 1. Introduction

Strategic decision-making at master-planned communities (MPCs) must incorporate intangible benefits when assessing large investments in amenity and infrastructure projects. Because such choices hinge on factors that reflect residents' *subjective* measures of quality of life, affordability, and risk, *objective* ROI analyses employed ubiquitously at profit-driven enterprises are often inadequate as a framework for making these important decisions.

To help MPCs make better choices for the future – and democratize decision-making – we developed a strategic planning tool, referred to herein as "Strategizer", that encapsulates community members' subjective preferences. The tool is intended to bring these preferences to bear in the decision-making process itself in the form of data-driven recommendations to committees, boards of directors, and executives tasked with making final decisions on funding costly projects. Using concepts drawn from utility theory [1], Strategizer prioritizes alternative amenity and infrastructure proposals based on precise and easy-to-collect measurements of residents' anticipated utilization, cost sensitivity, and risk tolerance.

The inspiration for this paper came from long-range strategic planning efforts at Tellico Village, an active adult lifestyle community in East Tennessee. Early testing of the concepts and methods presented here, as well as survey response data used in the decision-making examples, was facilitated by the support of members and associates of Tellico Village's Long-Range Planning Advisory Committee.

## 2. Decision Types

The discussion throughout this paper relates to amenity plans unless decision-making for infrastructure plans is mentioned explicitly. Some examples of popular community amenities include pickleball courts, clubhouses, swimming pools, recreation centers, and community centers. Strategizer can be used in the early concept phase of planning to identify promising candidates, or later when final decisions need to be made on funding pre-qualified, fleshed-out plans.

Given two or more <u>amenity plans</u>, Strategizer:

  i. Ranks plans by considering residents' anticipated utilization and cost sensitivity to each amenity
  ii. Estimates the percentage of households that favor the preferred plan
  iii. Predicts whether residents would actually be willing to fund the preferred plan

In the Strategizer framework, each plan under consideration contains one or more amenity *attributes*, which are specific, costly assets with characteristics or capabilities that distinguish them from other attributes in the plan. The general use case involves ranking multiple plans each comprising "bundles" of different amenity attributes (Figure 1).

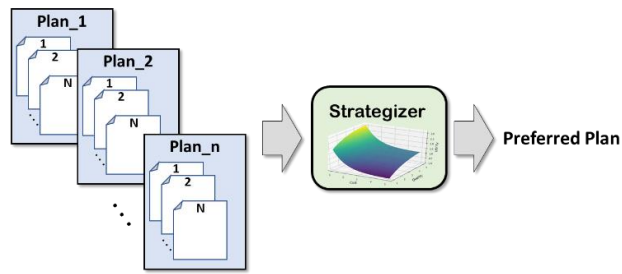

Figure 1. Strategizer selects the preferred plan by summing the utilities of all attributes in each plan and comparing expected utilities across the plans.

Typically, each attribute has both a "cost" component and a "quality" component, explained in detail in the following sections, although the former instead can be associated with just the bundle as a whole.

## 3. Utility of Quality

The quality of an attribute q reflects the collective quality of experiences the introduction of a new asset can bring to residents in a community. For example, the attribute "new outdoor swimming pool complex" enhances the quality of outdoor recreational experiences. If the project is funded and the pool is constructed, this quality will be greater than the reference quality, i.e., the level of quality $q_{ref}$ that existed prior to completion of the project.

Quality Attributes can be broken down into specific factors such as durability, aesthetics, and functionality. For instance, a swimming pool constructed with high-quality, durable materials will not only look appealing (aesthetics) but also require less maintenance over time (durability), enhancing the overall experience for residents. Features like a heating system or accessibility options add to the functionality, further improving the quality of the amenity.

From the standpoint of utility theory, we can assume (notwithstanding a highly unlikely lemon or eye-sore outcome) there is no downside quality risk for community plans and that completed projects always improve residents' quality of experiences and therefore increase their level of satisfaction, i.e., the quality component of utility U. To model how utility increases with quality, Strategizer uses a unit exponential function U(q) whose marginal utility diminishes as quality increases. The convergence constant has the same nominal value for all plans, ensuring consistency in how quality impacts utility across different projects.

Because there is no risk of losing quality of experiences when funding community projects, we make the assumption that slight variations across plans in the convergence rate from a common reference quality level $q_{ref}$ to higher quality levels can be ignored as long as we arrive at a better means of comparison. To do this, we assume that a community's average anticipated utilization $\bar{Q}$ of an asset ("usage frequency" also referred to as "consumption frequency", "usage rate", or "quantity demanded") is a good predictor of actual usage of the asset and can therefore be an effective proxy for quality of experience – more effective than direct measures of "preference". In econometric terms, usage frequency is an important variable that captures the expected consumption frequency or utilization rate of the amenity.

A Strategizer questionnaire utilizes a discrete quality scale from lower limit L (very low expected use) to upper limit H (very high expected use). The unit utility of quality function varying from 0 to 1 utils is scaled by the corresponding weight W, which lies between 1 and the maximum quality scaling factor $W_q$ (Figure 2):

1) $W(\bar{Q}) = \frac{(W_q-1)(\bar{Q}-L)+H-L}{H-L}$

In the example of Figure 3, the utility function of the attribute is valid over the domain $L = 1$ and $H = 5$, and average anticipated utilization $\bar{Q}$ = 2.6. Using $W_q$ = 2, the satisfaction level as measured in utils is obtained by multiplying the corresponding unit utility function by W = 1.4.

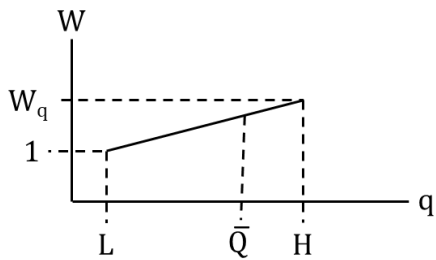

Figure 2. Diagram of scaling function for calculating quality weights.

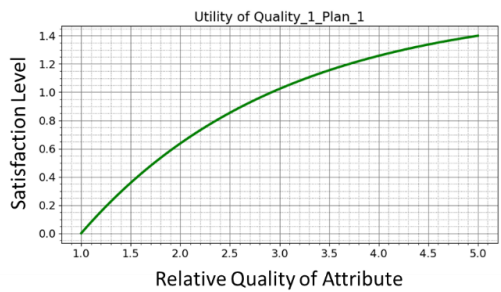

Figure 3. The unit utility of quality function is scaled by $W = 1.4$. Utility increases for a community as the quality of experiences derived from introducing an amenity asset increases.

## 4. Utility of Cost

The cost of an attribute c reflects the cost needed to introduce a new asset into the community and thereby attain a collective quality of experience q. Because a community's satisfaction level decreases with higher costs, utility of cost decreases with increasing cost. For example, residents will be very satisfied if the cost of the outdoor swimming pool complex is very low ($c = L$) and very unsatisfied if its cost is very high ($c = H$).

To model utility of cost, Strategizer begins with a unit exponential function $U(c)$ varying from 1 to 0 (Figure 4). In practice this can be achieved by applying the linear transformation:

2) $\quad T : U \to 1 - U$

to the unit utility of quality function and using a different exponential constant. The transformed function decreases steeply at low cost values, thereby ensuring that, for a given cost, the utility of cost-sensitive residents is always less than that for less cost-sensitive residents. The exponential convergence constant for cost, also referred to below as the cost tolerance constant, is itself indicative of cost tolerance, with larger values of the constant leading to higher, flatter utility of cost curves.

The standard gamble [2,3] has been accepted as the "gold standard" for the elicitation of expected utility when risk or uncertainty is involved in decisions. In this application, the cost of an amenity attribute is itself the independent variable subject to risk. In principle, a community's cost sensitivity to an amenity can be measured and translated into a convergence constant for the utility of cost function by presenting a standard gamble scenario to questionnaire respondents, as depicted in Figure 5. According to utility theory, given a choice between an extreme gamble and no gamble at all, there exists a probability $P_i$, referred to as the indifference probability, for which residents will be indifferent between the two alternatives. The two extreme outcomes in this situation are:

I. Only a slight increase in cost due to implementing the plan ($c = L$)
II. A substantial increase in cost due to implementing the plan ($c = H$)

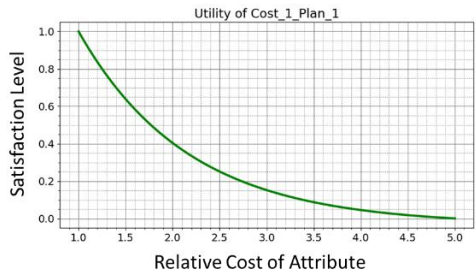
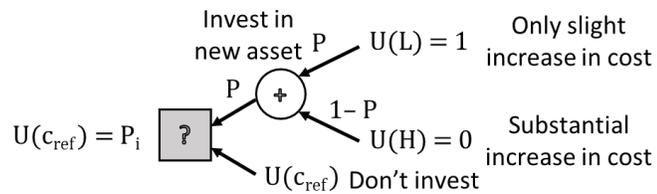

Figure 4. Utility of cost decreases as the cost of an amenity asset increases. The rate of decrease depends on the community's sensitivity to the project cost.

Figure 5. Gamble used in principle to measure the indifference probability of cost.

By rejecting the gamble, residents remain with the status quo, i.e., no investment in the amenity. We select a baseline reference cost $c_{ref}$ that represents the cost of the reference quality of experience $q_{ref}$ without investment in the new amenity. Any value between $L$ and $H$, subject to the constraint discussed in Section 9, can be assigned to represent this cost for the purpose of calculating the convergence constant.

The expected utility $U$ is the sum of the utilities of the extreme outcomes weighted by their probabilities, and $P$ is the probability of only a slight increase in cost:

3) $\quad U = P \cdot U(L) + (1 - P) \cdot U(H) = P$

since $U(L) = 1$ and $U(H) = 0$. Residents are indifferent between the outcome of the gamble and the status quo when the probability in Eq. 3 is the same as the utility of the reference cost:

4) $\quad P = U(c_{ref}) \equiv P_i$

Once the gamble is taken and $P_i$ is measured, the cost tolerance constant can be derived – and the utility of cost function constructed – based on three established points: ($c_{ref}$, $P_i$) and endpoints (L, 1) and (H, 0). The method for arriving at the constant given three points [4] is not unique to this paper and is summarized in the appendix.

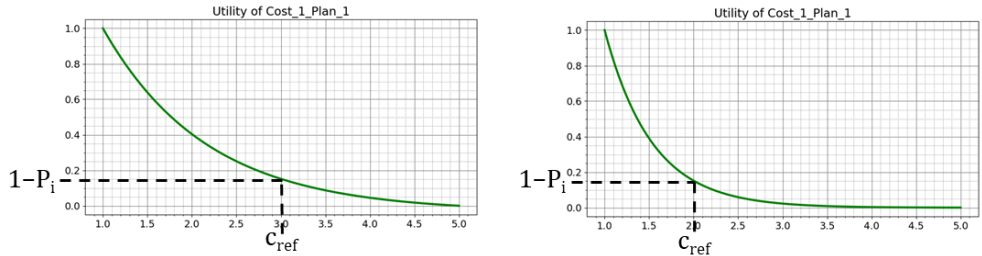

Figure 6. Formation of unit utility of cost. For a given indifference probability $P_i$, decreasing the reference cost $c_{ref}$ pulls down the utility curve, effectively decreasing cost tolerance across all plans. Recall that a downward sloping utility function undergoes a linear transformation so that larger values of measured $P_i$ lead to smaller convergence constants and steeper curves reflecting higher cost sensitivity.

It's evident from Figure 6 that for a fixed indifference probability, smaller values of the reference cost will lead to smaller values of the cost tolerance constant, and utility of cost will decline more steeply, analogous to the downward bending of a rubber band. This behavior is not an issue as long as the reference cost, which reflects residents' total existing costs without funding the new asset, is assigned the same value across all competing amenity plans. With this constraint in place, the indifference probabilities alone determine the convergence constants – and therefore the utility of cost functions – for each proposed amenity. Thereafter the reference cost and reference quality can be effectively shifted to the extreme left ($c_{ref} = q_{ref} = L$) for all plans, reflecting the reality that the gamble has been accepted and the amenity could achieve some satisfaction level at any point in the domain.

## 5. Measuring Cost Sensitivity

Initial trial runs of the survey methodology revealed that it would not be feasible to use the standard gamble method described above to measure indifference probabilities. This was because the method did not scale well; to obtain accurate results, survey administrators were needed to provide dedicated in-person assistance to describe the scenario to respondents and answer their questions, which proved too costly and time-consuming given that hundreds of measurement samples are needed to arrive at a statistically viable estimate of a community's cost sensitivity to a plan (the appendix provides more insight into sample size requirements).

To overcome the issues related to directly measuring indifference probabilities, a separate one-time survey was administered to a small sample of a resident population with the purpose of formulating a mathematical relationship between indifference probability and an easy-to-measure proxy for it. The survey questionnaire contained descriptions of a variety of realistic plans, including conceptual, well-defined, high-cost, medium-cost, unbundled, and bundled plans with multiple amenities. The cost of an attribute was framed as a hypothetical monthly cost that the respondent would incur for five years if the plan were to proceed. Two responses per plan were obtained:

i. Direct measurement of the indifference probability obtained by presenting an extreme gamble
ii. Maximum monthly cost the respondent would be willing to incur if the plan were to proceed

A second-order polynomial trend line was calculated from the dataset (Figure 7), indicating a coefficient of determination of $R^2$ = 0.822, which is significant enough for maximum monthly cost to be an effective proxy for direct measurement of the indifference probability, as long as it's applied uniformly across all plans being analyzed. Moreover, comparison of the polynomial trend line with a straight line approximation (Figure 8) revealed $R^2$ = 0.997, which justifies the use of a simple straight line approximation for converting maximum cost to indifference probability:

5) $P_i = 1 - \left(\dfrac{\text{maximum cost}}{\text{maximum possible cost}}\right)$

where maximum possible cost is the highest possible cost presented to respondents ($35 per month in the example).

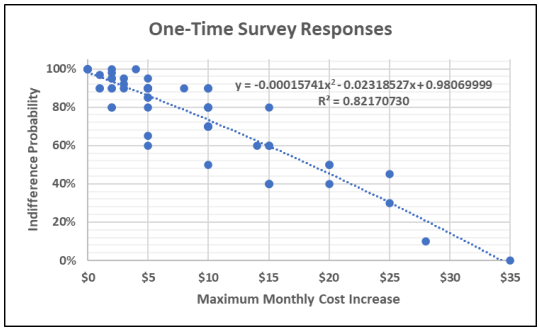

Figure 7. Relationship between indifference probability and maximum monthly cost respondents are willing to incur to approve an amenity plan.

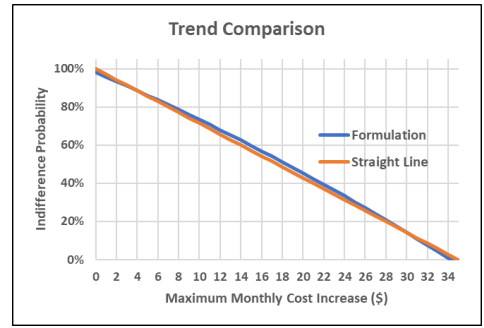

Figure 8. A straight line approximation can be used to translate maximum monthly costs into indifference probabilities.

## 6. Scaling Utility Components

Once the unit utility functions for cost and quality are derived, they are multiplied by their corresponding scaling factors and summed. The cost scaling factor is $W_c$, which typically has the same value as the maximum quality scaling factor $W_q$. The quality scaling factor $W(\bar{Q})$ is given by Eq. 1. The total utility of the attribute is the sum of the scaled quality and cost utility functions:

6) $U(c, q) = W_c \cdot U(c) + W(\bar{Q}) \cdot U(q)$

where $U(c)$ and $U(q)$ are the unit cost and unit quality utility functions.

## 7. Quality and Cost Scenarios

Once the utility function of an attribute is defined, a point on its surface must be selected for comparison with other attributes. For this purpose, Strategizer uses low, nominal and high cost-quality pairs or "targets" that reflect different amenity outcome scenarios:

   i. *Scenario A: Lowest cost, lowest quality targets (*$c_A$, $q_A$*).* The realization of the amenity achieves a smaller increase in quality over the reference quality compared with the other scenarios. The corresponding cost is also lower.

ii. *Scenario B: Nominal cost, nominal quality targets (*$c_B$, $q_B$*).* The realization of the amenity achieves a nominal increase in quality over the reference quality compared with the other scenarios. The corresponding cost is in the middle range compared with the other scenarios.

iii. *Scenario C: Highest cost, highest quality targets (*$c_C$, $q_C$*).* The realization of the amenity achieves a higher increase in quality over the reference quality compared with the other scenarios. The corresponding cost is also higher.

Between the time the survey questionnaire describing the plan is administered and the time the amenity is realized, there could be many alterations that might add to or detract from the original vision of the plan due to changes in the plan itself, project constraints, budgetary considerations, etc. Therefore, each scenario above represents a possible event with an associated probability that can be estimated by the plan architects or by Strategizer. The tool estimates each scenario probability by dividing the scenario's quality added by the quality added by all the scenarios. Quality added for a given scenario is the difference between the utility of its quality target and the utility of the previous quality target (or reference quality when calculating the probability of Scenario A).

This approach assumes that (a) each probability is proportional to the utility of quality its quality target adds to a plan relative to the preceding quality target (or reference quality target), (b) cost outcomes are positively correlated with quality outcomes, (c) the minimum amenity set (Scenario A) adds more to quality than the nominal amenity set (Scenario B), and so on, and (d) all the probabilities sum to one.

Strategizer utilizes a decision tree to calculate the expected utility of each plan for comparison across plans, as illustrated in Figure 9. For each low, nominal and high scenario, the utilities of all the attributes in a plan are calculated at their corresponding scenario cost and quality targets and added together. The expected utility of each plan is the sum of these terms weighted by the scenario probabilities $P_A$, $P_B$, $P_C$. The amenity plan with the highest expected utility is chosen.

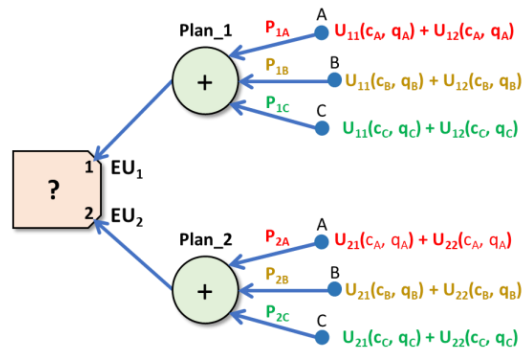

Figure 9. Comparison of two plans each having two attributes.
$U_{ij}$ = Utility of attribute j in Plan_i.

## 8. Decision-Making Example

Strategizer processes all the input data from a single Excel worksheet containing multiple tabs, one tab for each alternative plan. To perform an analysis, two sets of data are required for each attribute:

i. External data originating from the residents survey – includes all measured values of maximum cost and anticipated utilization

ii. Internal data provided by the decision-making team itself – includes scenario cost and quality targets specific to each attribute

The tool calculates the mean and standard deviation of the measurements for maximum cost and anticipated utilization and, together with default reference values for cost and quality, calculates the expected utilities for each plan using different scenarios for cost and quality targets as described in the previous section.

Figure 10 summarizes the data for comparing two amenity plans each having a single attribute. Notice that maximum cost for Plan_2 is greater than that for Plan_1, implying respondents on average prefer Plan_2 over Plan_1 from the standpoint of cost sensitivity. However, anticipated utilization is less for Plan_2, so it's not clear from this data alone which plan is preferred. Moreover, the plan administrators believe that the target quality and cost values will be higher for Plan_2 across all scenarios. Strategizer's comparison of expected utility between the two plans indicates Plan_1 is the preferred plan.

|  | Plan_1 | | | Plan_2 | | |  | Plan_1 | | | Status Quo | | |
|---|---|---|---|---|---|---|---|---|---|---|---|---|---|
|  | Max Cost ($0 - $35) | Utilization (1 - 5) |  | Max Cost ($0 - $35) | Utilization (1 - 5) |  |  | Max Cost ($0 - $35) | Utilization (1 - 5) |  | Max Cost ($0 - $35) | Utilization (1 - 5) |  |
| Mean | $5.33 | 2.5 |  | $5.83 | 2.33 |  | Mean | $5.33 | 2.5 |  | $5.33 | 2.5 |  |
| Stdev.S | $4.56 | 1.24 |  | $7.31 | 1.37 |  | Stdev.S | $4.56 | 1.24 |  | $4.56 | 1.24 |  |
|  | Cost | Quality | Probability | Cost | Quality | Probability |  | Cost | Quality | Probability | Cost | Quality | Probability |
| Low Target | 2 | 2 | 50% | 2.5 | 2.5 | 64% | Low Target | 2 | 2 | 50% | 1 | 1 | 50% |
| Med Target | 3 | 3 | 32% | 3.5 | 3.5 | 22% | Med Target | 3 | 3 | 32% | 1 | 1 | 32% |
| High Target | 4 | 4 | 18% | 4.5 | 4.5 | 14% | High Target | 4 | 4 | 18% | 1 | 1 | 18% |
| Expected Utility |  | 1.375 |  |  | 1.335 |  | Expected Utility |  | 1.375 |  |  | 2.000 |  |

Figure 10. Data used to decide between two plans. Only the mean and standard deviation of the measurements are shown. Strategizer selects Plan_1 because its expected utility is higher.

Figure 11. Go/no-go analysis of Plan_1 assuming cost scaling factor $W_c$ = 2 indicates a higher expected utility for the status quo, implying respondents are not in favor of funding the plan.

It is possible to make a reasonable guestimate as to whether a community would be willing to pay for a preferred plan. Figure 11 compares Plan_1 with the status quo, which has identical cost sensitivity and utilization numbers but with scenario targets set to the reference value (the extreme minimum $L = 1$ once the cost tolerance constant has been calculated). Since the expected utility is lower for the new amenity, Strategizer's recommendation is not to proceed with the project.

It was noted in Section 6 that the cost scaling factor $W_c$ is usually assigned the same value as the maximum quality scaling factor $W_q$. This represents the most cost-sensitive attitude toward funding a project. Scaling the parameter down decreases the cost sensitivity of the decision. For example, setting $W_c$ to the minimum of 1 decreases Plan_1's expected value from 1.375 to 1.115, and the status quo's expected value from 2.000 to 1.000, resulting in a "Go" decision to fund the plan.

## 9. Sensitivity Analysis

The Strategizer tool makes use of a robust set of decision tree libraries and utilities developed for making business decisions. Two capabilities designed for analysis of decision trees are especially relevant for amenity ranking and are highlighted in this section.

The first of these capabilities is sensitivity analysis. Strategizer can perform a variety of sweeps each having dozens of different scenario probability combinations to determine if the plan selection is sensitive to the choice of probabilities. The following log output shows the first two outcomes of a probability sweep using 2% for the increment parameter. In addition to expected utility, sensitivity analysis reports the "best decision" using six other established decision criteria.

Probability Sweep Results

Sweep 1

Result: 1 Probabilities: [ 44%  30%  26%  44%  30%  26%]
Option 1 is probably the best decision. Expected cost: 1.071
Expected cost of Option 2: 0.992  Difference: 0.079

Maximin criterion cost: 1.0 (Option 1)
Maximax criterion cost: 1.1 (Option 1)
Minimax regret criterion cost: -0.0 (Option 1)
Most likelihood criterion cost: 1.0 (Option 1)
Hurwicz criterion cost: 1.1 (Option 1)

Result: 2  Probabilities: [ 46%  28%  26%  46%  28%  26%]
Option 1 is probably the best decision. Expected cost: 1.07
Expected cost of Option 2: 0.991  Difference: 0.08

Maximin criterion cost: 1.0 (Option 1)
Maximax criterion cost: 1.1 (Option 1)
Minimax regret criterion cost: -0.0 (Option 1)
Most likelihood criterion cost: 1.0 (Option 1)
Hurwicz criterion cost: 1.1 (Option 1)

…

The second capability is the use of Monte Carlo simulations to predict preferences across the community population. The inputs of the analysis – the indifference probabilities $P_i$ and the quality weights $W(\bar{Q})$ – are turned into random variables (with mean and standard deviation calculated from the aggregate response data) and propagated through the system many times. The result is a histogram representing the difference between the expected utility of two plans. The example of Figure 12 indicates that out of 5,400 households, about 63% are expected to approve Plan_2 while 37% are expected to approve Plan_1. Monte Carlo analysis can also be performed for go/no-go type decisions.

To generate meaningful results for Monte Carlo simulations, it's important to properly constrain the indifference probability when it takes on random values. The extreme gamble presented in Section 4 highlights the fact that a rational investor favors the gamble only if the indifference probability is greater than the utility of the reference cost:

7) $P_i > U(c_{ref})$

Applying the linear transformation of Eq. 2 to both sides gives:

8) $1 - P_i < 1 - U(c_{ref})$

Figure 13 depicts a utility of cost function U that represents the right side of Eq. 8 with a convergence constant approaching infinity. The equation for the downward sloping line is given by:

9) $U(c) = \frac{H - c_{ref}}{H - L}$

Notice that $1 - P_i$ must always be under the line. Substituting the expression in Eq. 9 for the right hand side of the inequality in Eq. 8 leads to the following constraint on the indifference probability:

10) $P_i > \frac{c_{ref} - L}{H - L}$

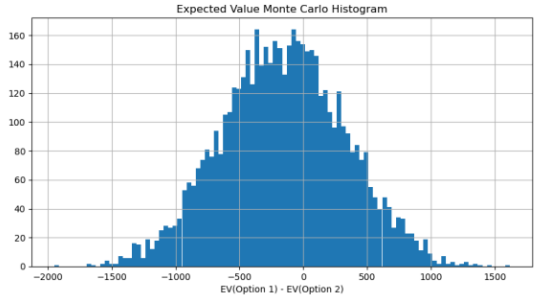

Figure 12. Monte Carlo simulation statistics comparing two plans: Count = 5400, Mean = -172, Standard Deviation = 483, Population Below Zero = 63.4%.

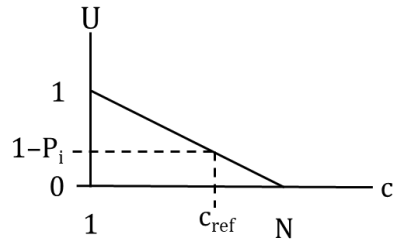

Figure 13. Utility of cost with infinite convergence.

## 10. Infrastructure Decisions

A significant portion of decisions in large communities are those that involve investments in costly infrastructure projects. Examples include upgrade of water distribution and sewer systems, road re-pavement, building renovation or replacement, and trash/solid waste removal. Unlike amenities, infrastructure projects are usually unavoidable for the long term sustainability of the community, and decisions come down to selecting among alternative bids submitted by public works contractors. Decision makers often rely on community members or outside consultants who have the appropriate domain expertise to understand the issues, arrive at an infrastructure plan, communicate the plan requirements to contractors, and make a final recommendation on which contractor proposal is most closely aligned to the needs of the community.

A proposal or bid can be assumed to be viable if its cost is commensurate with the extent to which it will likely mitigate the risk of catastrophic failure over many years. From the perspective of decision-making, a low cost, low risk mitigation proposal is as equally viable as a high cost, high risk mitigation proposal, and decision makers must determine which cost-risk combination best reflects the preferences of the community.

Given a specific infrastructure plan under consideration, Strategizer determines which type of plan implementation respondents prefer based on residents' risk and cost sensitivities:

- A low cost, low risk mitigation alternative
- A high cost, high risk mitigation alternative

Knowing which type of implementation residents prefer will aid in deciding which bid to select.

The survey questionnaire does not ask residents to evaluate detailed bids. Instead, it includes a basic description of a single infrastructure plan, explaining why it's needed and what needs to be done. As with the amenity surveys, cost sensitivity is measured by presenting respondents a range of monthly costs and asking them to select the maximum cost they would be willing to incur. Risk sensitivity is measured in a similar manner, with operational lifespan serving as proxy for risk mitigation (Figure 14).

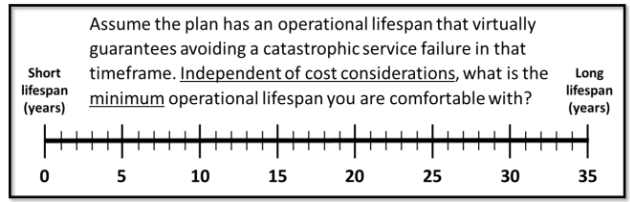

Figure 14. Survey question designed to calculate the indifference probability of risk for an infrastructure plan.

Strategizer translates the maximum cost and minimum lifespan numbers into indifference probabilities to compute the exponential convergent constants for cost tolerance C and risk tolerance R. The preferred plan can be selected simply by comparing the cost/risk tolerance constants:

- If $R < C$ (lower tolerance to risk than cost), respondents prefer high cost, high risk mitigation
- If $C < R$ (lower tolerance to cost than risk), respondents prefer low cost, low risk mitigation

Strategizer can generate the utility functions for cost and risk to compare different cost-risk options, and perform sensitivity analysis and Monte Carlo simulations as described in Section 9. Rather than comparing two separate plans as with amenity analysis, the same plan is compared with itself using different scenarios for cost and risk targets: one for low cost, low risk mitigation and the other for high cost, high risk mitigation.

## 11. Conclusion

The journey to democratize strategic decision-making in master-planned communities is not without its challenges, but the tools and methodologies presented in this paper and our previous work on forecasting housing growth dynamics in MPCs [5] together offer important steps toward achieving this vision. Strategizer, the strategic planning tool introduced here, is specifically designed to quantify residents' subjective preferences about major investments in amenities and infrastructure projects, transforming these preferences into actionable insights. By developing accessible and data-driven frameworks, we aim to empower residents, planners, and stakeholders alike to actively engage in shaping the future of their communities. This inclusive approach promotes collaboration, giving voice to individuals who have historically been underrepresented in strategic planning processes.

Our approach shifts strategic planning from a top-down exercise to a more open, participatory model, ensuring that diverse perspectives are incorporated into key decisions. Drawing on data from brief online surveys, Strategizer ranks alternative plans by considering anticipated utilization and cost or risk sensitivity, thereby providing data-driven recommendations to decision makers. By making the underlying data and insights accessible to all stakeholders, we foster a culture of transparency and trust, which is crucial to long-term community well-being and resilience.

Ultimately, Strategizer is more than just a technical solution—it represents a commitment to community empowerment. It aligns with established techniques such as Contingent Valuation, Stated Preference, Conjoint Analysis, and Social Exchange Theory [6,7,8], while minimizing cognitive and time burdens on survey respondents. By democratizing the strategic planning process, we hope to inspire a shared vision where every resident has the opportunity to contribute meaningfully to the community's growth and development. This paper lays the foundation for such an evolution, advocating for a more inclusive, adaptive, and resident-centric approach to planning.

## 12. Appendix

### Utility Function

The utility function used for Strategizer analysis is an exponential of the form:

11) $U(x) = a - be^{-x/K}$

where K is a convergence constant. If L and H are the domain endpoints, then for an increasing unit exponential function:

12) $U(L) = a - be^{-\frac{L}{K}} = 0$

13) $U(H) = a - be^{-\frac{H}{K}} = 1$

Solving for two equations in two unknowns gives:

14) $a = \frac{e^{-L/K}}{e^{-L/K} - e^{-H/K}}$

15) $b = \frac{1}{e^{-L/K} - e^{-H/K}}$

If $c_{ref}$ is the reference value for quality or cost, then $U(c_{ref}) = P_i$ from Eq. 4 and K can be solved by incrementing it until $\left|a - be^{-\frac{c_{ref}}{K}} - P_i\right| <$ a small error. Once the convergence constant is determined, Eq. 11 can be used for unit utility of quality, and the unit utility of cost is obtained using the linear transformation in Eq. 2. A similar procedure is used to determine the convergence constant for risk and its corresponding utility function.

### Sample Size Requirements

The t-distribution of the response data obtained from an initial dry run of the methodology was used to determine the minimum sample size N needed for measuring maximum cost and anticipated utilization of amenity plans:

16) $N = 4\left(\frac{t_{n-1} s}{w}\right)^2$

where: s = estimate of the population standard deviation, w = required width of the confidence interval and n = number of samples. There are n – 1 degrees of freedom.

Population standard deviation was estimated assuming almost all samples are within six standard deviations of the extreme limits, and that 95% confidence ($\alpha$ = 5%) is required. The estimates are:

- Maximum cost: s = (\$30 – \$0)/6 = \$5; w = \$1.00 → N = 485
- Anticipated utilization: s = (5 – 1)/6 = 2/3; w = 0.25 → N = 138

These calculations imply that maximum cost is the most stringent constraint on sample size. For this measurement, one can be 95% confident that the average obtained is within \$0.50 of the true population average – whatever that is – assuming the vast majority of responses come in between \$0 and \$30.